\documentclass[12pt,letterpaper]{article}
\usepackage{amsmath,amssymb}
\usepackage{epstopdf}
\usepackage{graphicx}
\usepackage{hyperref}
\usepackage{booktabs}
\usepackage{fullpage}

\newcommand\given{{\:|\:}}
\newcommand\conea{{\text{C1a}}}
\newcommand\coneb{{\text{C1b}}}
\newcommand\ctwo{{\text{C2}}}
\newcommand\cthree{{\text{C3}}}
\newcommand\cfour{{\text{C4}}}
\newcommand\oone{{\text{O1}}}
\newcommand\otwo{{\text{O2}}}

\newcommand\cfive{{\text{C5}}}

\newcommand{\pjt}[1]{\textcolor{black}{#1}}

\usepackage{xcolor}

\newcommand*\rot{\rotatebox{90}}

\begin{document}

\begin{flushleft}
{\Large
\textbf\newline{A factor graph EM algorithm for inference of kinetic microstates from patch clamp measurements} 
}
\newline
\\
Alexander S. Moffett\textsuperscript{1},
Guiying Cui\textsuperscript{2},
Peter J. Thomas\textsuperscript{3},
William D. Hunt\textsuperscript{4},
Nael A. McCarty\textsuperscript{2},
Ryan S. Westafer\textsuperscript{5},
Andrew W. Eckford\textsuperscript{1,*}
\\
\bigskip
\textbf{1} Dept. of Electrical Engineering and Computer Science, York University, 4700 Keele Street, Toronto, ON M3J 1P3, Canada
\\
\textbf{2} Emory + Children’s Center for Cystic Fibrosis and Airways Disease Research, Emory University School of Medicine and Children’s Healthcare of Atlanta, Atlanta, GA 30322, USA
\\
\textbf{3} Dept. of Mathematics, Applied Mathematics, and Statistics, Case Western Reserve University, 10900 Euclid Avenue, Cleveland, OH 44106, USA
\\
\textbf{4} School of Electrical and Computer Engineering, Georgia Institute of Technology, 777 Atlantic Drive NW, Atlanta, GA 30332, USA
\\
\textbf{5} Georgia Tech Research Institute, 400 10th St NW, Atlanta, GA 30318, USA
\bigskip

%
%





* Corresponding author: aeckford@yorku.ca

\end{flushleft}





\section*{Abstract}
We derive a factor graph EM (FGEM) algorithm, a technique that permits combined parameter estimation and statistical inference, to determine hidden kinetic microstates from patch clamp measurements. 
Using the cystic fibrosis transmembrane conductance regulator (CFTR) and nicotinic acetylcholine receptor (nAChR) as examples, we perform {\em Monte Carlo} simulations to demonstrate the performance of the algorithm. We show that the performance, measured in terms of the probability of estimation error, approaches the theoretical performance limit of maximum {\em a posteriori} estimation. Moreover, the algorithm provides a reliability score for its estimates, and we demonstrate that the score can be used to further improve the performance of estimation.
We use the algorithm to estimate hidden kinetic states in lab-obtained CFTR single channel patch clamp traces.

%
%
%
%


\renewcommand{\baselinestretch}{1.5}
\normalsize

\section{Introduction}


The objective of Bayesian inference is to obtain the {\em a posteriori} probability $p(x\given y)$ of a hidden random variable $x$ given observations $y$. Many algorithms exist for calculating Bayesian inference, either exactly or approximately \cite{mackay2003}. In one important class of inference algorithms, the stochastic model is expressed on a graph, and the inference algorithm is performed by passing messages over this graph. An example of such a graphical representation is the factor graph, where the associated inference algorithm is the sum-product algorithm \cite{kschischang2001}. Applications of inference algorithms are found in diverse areas such as bioinformatics \cite{xiong2011,meyer2019simultaneous} and biophysics \cite{metzner2009estimating,potrzebowski2018bayesian}, telecommunications, and more recently in machine learning \cite{barber2012}.

The classic work of Colquhoun and Hawkes gave an algorithm to estimate transition rates among kinetic microstates, given only observable data such as whether an ion channel was open or closed \cite{colquhoun1981}. Related algorithms have long been known in the signal processing literature, such as the Baum-Welch algorithm \cite{baum1970}; these algorithms are special cases of the Expectation-Maximization (EM) algorithm \cite{dempster1977}. Since this early work on inference of kinetic models of ion channels from electrophysiological data, there has been an abundance of work on improving different aspects of analysis using distinct computational approaches. Several methods have been developed to automate idealization of noisy ion channel current recordings, including EM algorithm \cite{shah2018tracespecks} and deep learning \cite{celik2020deep} approaches. Non-parametric Bayesian approaches have been used to identify the number of kinetically distinct hidden states without the need to specify a model \cite{hines2015analyzing}. Finally, many methods have been developed for estimation of the hidden state transition matrix, including maximum likelihood methods \cite{qin1996estimating,colquhoun1996joint,colquhoun2003quality,moffatt2007estimation,nicolai2013solving} and Bayesian approaches \cite{rosales2001,gin2009markov,siekmann2011,epstein2016bayesian}.

In this article, we introduce a factor graph EM (FGEM) algorithm \cite{eckford2005,loeliger2007,dauwels2009} which is able to estimate the transition matrix while also yielding maximum likelihood hidden state trajectories from ion channel electrophysiology data. The main advantage of this method is its accuracy and efficiency in calculating \emph{a posteriori} probabilities, meaning the probabilities of hidden states given observations. Our FGEM algorithm thus provides parameter estimates while simultaneously furnishing insight into the hidden ion channel conformational dynamics, given a kinetic model. Although frequently applied to parameter estimation problems in digital signal processing, FGEM algorithms have recently been applied to biomedical signal processing problems \cite{wadehn2019model,wadehn2019state}, and has not hitherto been applied to patch clamp signals.

\begin{figure}[h!]
    \centering
    \includegraphics[width=3in]{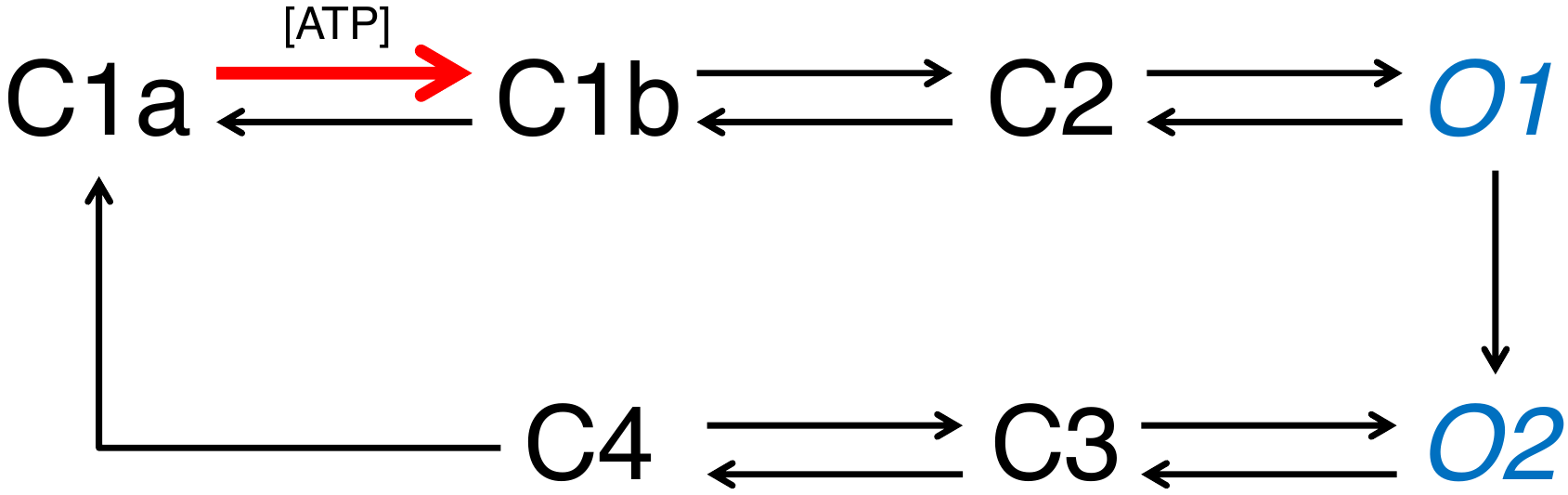}
    \vspace{2ex}
    \caption{CFTR cycle model. Permitted state transitions are indicated with arrows. State transitions that are sensitive to ATP concentration are indicated with bold red arrows and labelled [ATP]. States with open and closed ion channels are indicated in blue italics and black, respectively. See also reference \cite{fuller2005block} and equation (\ref{eqn:R-CFTR}).}
    \label{fig:CFTR_model}
\end{figure}

\begin{figure}[h!]
    \centering
    \includegraphics[width=2.3in]{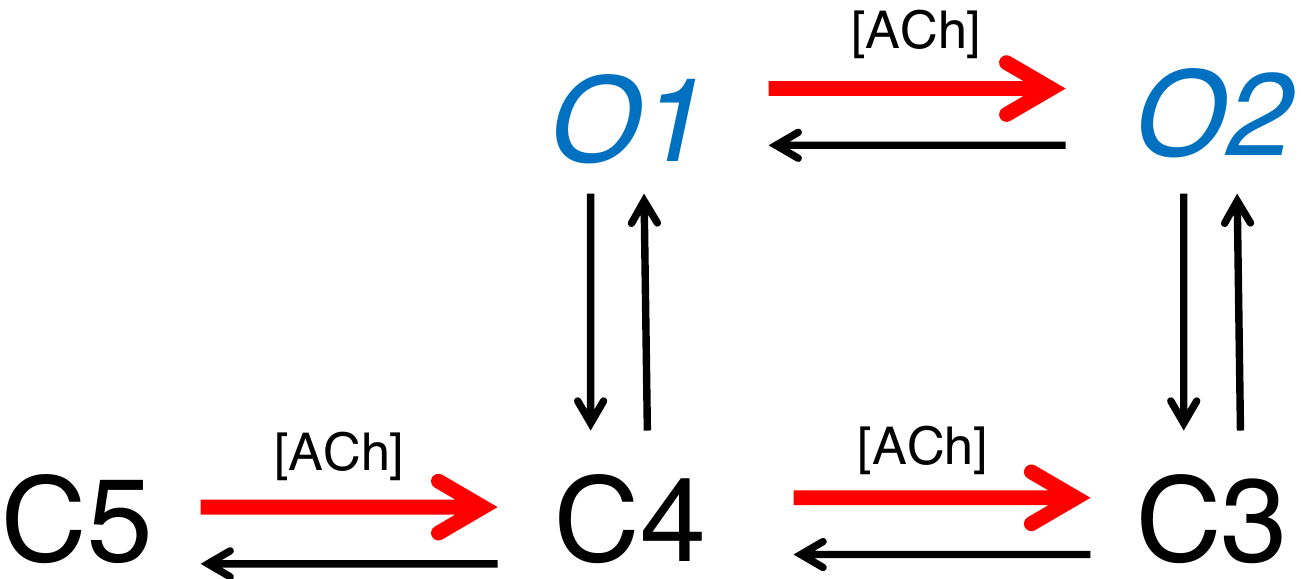}
    \vspace{2ex}
    \caption{nAChR cycle model. Permitted state transitions are indicated with arrows. State transitions that are sensitive to ACh concentration are indicated with bold red arrows and labelled [ACh]. States with open and closed ion channels are indicated in blue italics and black, respectively. See also references \cite{colquhoun1982,eckford2018channel} and equation (\ref{eqn:R-ACh}).}
    \label{fig:ACh_model}
\end{figure}

We apply our inference technique to the cystic fibrosis transmembrane conductance regulator (CFTR) anion channel and a nicotinic acetylcholine receptor (nAChR) \cite{csanady2019structure}. Cystic fibrosis (CF) is a life-threatening genetic disease affecting the respiratory and digestive systems, caused by mutations to CFTR \cite{rey2019cystic}. CFTR is a ``broken'' member of the ATP-binding cassette (ABC) transporter class, in that CFTR acts as an ATP-gated ion channel rather than an active transporter as is the function of other ABC transporters. CFTR consists of a single polypeptide chain, with two transmembrane domain (TMD)-nucleotide binding domain (NBD) pairs connected through a region called the R domain \cite{zhang2016atomic}. The two TMDs form a gated channel which is controlled by the state of the two intracellular NBDs. 
Each of the two NBDs contributes to two binding sites for ATP, although only one of these sites facilitates the hydrolysis of ATP to ADP. There are a number of kinetic models of the CFTR cycle, varying in both the number of microstates and the balance between reversible and irreversible transitions \cite{vergani2003mechanism,fuller2005block,csanady2010strict}. In any model, the steps coupled to ATP hydrolysis must be irreversible, while it can be argued that other transitions may be treated as effectively irreversible \cite{csanady2010strict}. In this work, we use a modified version of the model from Fuller \emph{et al.} \cite{fuller2005block} (Fig.~\ref{fig:CFTR_model}). 

nAChRs are ligand-gated ion channels activated by the neurotransmitter acetylcholine (ACh) \cite{albuquerque2009mammalian}. nAChRs are composed of five monomers of varying compositions. These subunits are arranged around a central pore, through which ions can flow. There are multiple ACh binding sites, and ACh binding to these sites induces conformational changes in the nAChR which open the channel to Na$^{+}$, K$^{+}$, and Ca$^{2+}$ ions \cite{unwin2013nicotinic}. Several kinetic models of the muscular nAChR have been developed \cite{colquhoun1982,edelstein1996kinetic,carignano2016analysis}, accounting for the allosteric nature of ACh-induced nAChR opening \cite{changeux2018nicotinic}.
\pjt{We use a version of the model adapted from \cite{colquhoun1982} (Fig.~\ref{fig:ACh_model}).}

We demonstrate the capability of the FGEM algorithm to infer hidden state trajectories with low probability of error for simulated CFTR and nAChR data with varying ligand concentrations and for experimental CFTR current recordings.

\section{Materials and methods}

\subsection{Receptor model}







We consider the CFTR and nAChR membrane channels.  Both have associated receptor elements---CFTR has receptor regions for ATP and nACHR contains receptors to ACh.  These can be modelled using a master equation of the form
\begin{align}
    \label{eqn:master}
    \frac{dY}{dt} = YR ,
\end{align}
where $Y$ is a row vector with length equal to the number of hidden kinetic states, and $R$ is a square matrix of kinetic rates for each possible state transition. In this formulation, $Y_i$ is the fraction of receptors in kinetic state $i$, while $R_{ij}$ is the transition rate from state $i$ to state $j$. In some kinetic states, an ion channel is open, allowing a current to flow; in other states, the channel is closed, allowing no current. A patch clamp measures the current flowing through the channel, indicating whether the membrane channel is open or closed.

For CFTR, we use a purely cyclical model of the ATP-driven CFTR gating cycle, based on the model in \cite{fuller2005block}, as shown in Figure \ref{fig:CFTR_model}. In this 7-state model, states \conea{}, \coneb{}, and \cfour{} are fully closed, so we assume that ions are completely unable to pass through when CFTR is in these states. States \ctwo{} and \cthree{} are closed-permissive states\pjt{; we assume here that they also pass no current.} 
States  \oone{} and \otwo{} are open states in which chloride can flow through CFTR. Beginning with \conea{}, with a single ATP bound at the first ATP binding site, the reversible transition to \coneb{} occurs when a second ATP binds so that both binding sites are occupied. 
Because this step involves ATP binding, the rate depends on the concentration of ATP. The reversible transitions from \coneb{} to \ctwo{} and from \ctwo{} to \oone{} are conformational changes resulting in an open CFTR pore. The transition from \oone{} to \otwo{} is the first of two irreversible steps in the cycle, with one NBD-bound ATP undergoing hydrolysis to ADP. CFTR can then undergo reversible conformational changes from \otwo{} to \cthree{} and from \cthree{} to \cfour, resulting in a closed pore. 
Finally, the second irreversible step occurs in the transition from \cfour{} to \conea{}, where the NBD-bound ADP unbinds from CFTR, leaving one apo and one filled ATP binding site. 

The transition rates are given in the rate matrix (with rows and columns, in order, corresponding to $\conea$, $\coneb$, $\ctwo$, $\oone$, $\otwo$, $\cthree$, and $\cfour$):
\begin{align}
    \label{eqn:R-CFTR}
    R &= 
    \left[
        \begin{array}{ccccccc}
            R_{11} & k_{\conea\rightarrow{}\coneb} & 0 & 0 & 0 & 0 & 0 \\
            k_{\coneb\rightarrow\conea} & R_{22} & k_{\coneb\rightarrow\ctwo} & 0 & 0 & 0 & 0 \\
            0 & k_{\ctwo\rightarrow\coneb} & R_{33} & k_{\ctwo\rightarrow\oone} & 0 & 0 & 0 \\
            0 & 0 & k_{\oone\rightarrow\ctwo} & R_{44} & k_{\oone\rightarrow\otwo} & 0 & 0 \\
            0 & 0 & 0 & 0 & R_{55} & k_{\otwo\rightarrow\cthree} & 0 \\
            0 & 0 & 0 & 0 & k_{\cthree\rightarrow\otwo} & R_{66} & k_{\cthree\rightarrow\cfour} \\
            k_{\cfour\rightarrow\conea} & 0 & 0 & 0 & 0 & k_{\cfour\rightarrow\cthree} & R_{77}
        \end{array}
    \right]
\end{align}
with the diagonal entries $R_{ii}$ set so that each row of $R$ sums to zero, and with $R_{ij} = 0$ indicating that the transition $i \rightarrow j$ is forbidden. For completeness, the full continuous-time master equation describing this system is given in the SI.

Table \ref{tab:CFTR-parameters} gives example values for each rate, noting that the value of $k_{\conea\rightarrow\coneb}$ is dependent on the environmental concentration of ATP. 

For nAChR, we adapted a 5-state model of the nAChR cycle from \cite{colquhoun1982} (Figure \ref{fig:ACh_model}). This model includes three closed states (\cthree{}, \cfour{}, and \cfive{}) and two open states (\oone{} and \otwo{}). The transitions from \cfive{} to \cfour{}, \cfour{} to \cthree{}, and \oone{} to \otwo{} correspond to binding of acetylcholine to the receptor.  These rates are dependent on acetylcholine concentration. 

The transition rates are given in the rate matrix (with rows and columns, in order, corresponding to $\oone$, $\otwo$, $\cthree$, $\cfour$, and $\cfive$):
\begin{align}
    \label{eqn:R-ACh}
    R &= 
    \left[
        \begin{array}{ccccc}
            R_{11} & k_{\oone\rightarrow{}\otwo} & 0 & k_{\oone\rightarrow\cfour} & 0 \\
            k_{\otwo\rightarrow\oone} & R_{22} & k_{\otwo\rightarrow\cthree} & 0 & 0 \\
            0 & k_{\cthree\rightarrow\otwo} & R_{33} & k_{\cthree\rightarrow\cfour} & 0 \\
            k_{\cfour\rightarrow\oone} & 0 & k_{\cfour\rightarrow\cthree} & R_{44} & k_{\cfour\rightarrow\cfive} \\
            0 & 0 & 0 & k_{\cfive\rightarrow\cfour} & R_{55}
        \end{array}
    \right]
\end{align}
with the same conditions on $R$ as in (\ref{eqn:R-CFTR}). Again, the full master equation is found in the SI.

Values for the rate parameters (taken from \cite{colquhoun1982}) are found in Table \ref{tab:ACh-parameters}, noting that  $k_{\oone\rightarrow\otwo}$, $k_{\cfive\rightarrow\cfour}$, and $k_{\cfour\rightarrow\cthree}$ are dependent on ACh concentration.

\subsection{Inference and patch clamp signals}

If all system parameters are known, the sum-product algorithm \cite{kschischang2001,Eckford2004Thesis} may be used to determine the {\em a posteriori} distribution of the kinetic microstates at any time, given the patch clamp observations. Previous approaches to this problem have involved histogram fitting \cite{csanady2000rapid}.

Formally, in the absence of noise, our system contains a set $\mathcal{Y}$ of observable states, a set $\mathcal{S}$ of kinetic microstates, and a mapping $m:\mathcal{S} \rightarrow \mathcal{Y}$ from microstates to observable states. Using CFTR as an example, we have
\begin{align}
    \label{eqn:Yset}
    \mathcal{Y} &= \{ 0, 1 \} \\
    \label{eqn:Sset}
    \mathcal{S} &= \{ \conea, \coneb, \ctwo, \cthree, \cfour, \oone, \otwo\} \\
    m(s) &= 
        \left\{ 
            \begin{array}{cl} 
                0, & s \in \{ \conea, \coneb, \ctwo, \cthree, \cfour \} \\
                1, & s \in \{ \oone, \otwo \} 
            \end{array}
        \right.
\end{align}
The observable states $\{ 0,1 \}$ correspond to the ion channel being closed and open, respectively. The microstates are labelled so that the first letter indicates whether the channel is closed (C) or open (O).

Patch clamp signals are normally sampled in time to form discrete-time signals. Let $y = [y_1,y_2,\ldots,y_n] \in \mathcal{Y}^n$ represent the sequence of observations for a single channel, and let $s = [s_1,s_2,\ldots,s_n] \in \mathcal{S}^n$ represent the corresponding microstates (i.e., for each $k$, $y_k = m(s_k)$). The transitions of microstates $s_{k-1} \rightarrow s_k$ can be modelled as Markov chains: let $P = [P_{ij}]$ represent a $|\mathcal{S}|\times|\mathcal{S}|$ transition probability matrix, with $P_{ij} = \Pr(s_k = j \given s_{k-1} = i)$. (We may also allow $P$ to be time-varying; in this case we will write the matrix at time $k$ as $P^{(k)} = [P_{ij}^{(k)}]$.)

The stochastic dynamics of the microstates are normally stated in terms of a rate matrix $R = [R_{ij}]$, where $R_{ij}$ is the kinetic rate of the transition from $i$ to $j$. Given $R$, and a sufficiently small discrete time step $\Delta t$, $P$ is given by
%
\begin{align}
    \label{eqn:P}
    P = I + R \Delta t 
    + o(\Delta t),
\end{align}
where $I$ is the identity matrix of the same size as $R$, and $\lim_{h\to 0}(o(h)/h)=0$. The master equation is thus converted to a discrete-time Markov chain \cite{smith2002modeling}.

Given this system, we use the sum-product algorithm to obtain the {\em a posteriori} distribution $p(s_k \given y)$, the complete details of which are described in the supplemental information (SI). Given $p(s_k \given y)$, we can then estimate the hidden state by finding the most probable state $s_k$ given $y$, i.e.,
\begin{align}
    \hat{s}_k &= \arg \max_{s_k \in \mathcal{S}} p(s_k \given y) ,
\end{align}
which is known as the maximum {\em a posteriori} probability (MAP) estimate. It can be shown that the MAP estimate has the lowest probability of error of any estimator.


\subsection{The factor graph EM algorithm}

The transition probability matrix $P$ (from (\ref{eqn:P})) is unknown {\em a priori} and must be estimated from the data.  This estimation task must occur alongside the inference of the hidden states. This is a natural setting for the EM algorithm \cite{dempster1977}. 

In our application of the EM algorithm, we treat the \pjt{nonzero entries of the} transition probability matrix $P$ (and potentially the observation noise variance $\sigma^2$) as unknown parameters. 
Starting with initial guesses of $P$ (and $\sigma^2$), the EM algorithm proceeds iteratively.
First we perform inference, as though the parameter estimates were equal to their true values (E-step), and then maximize the log of the inferred likelihood function to obtain new estimates of $P$ (and $\sigma^2$) (M-step). Subsequently these estimates are fed back to the E-step to complete a new iteration. 

The factor graph EM algorithm is a variant of the EM algorithm that is intended for use alongside the sum-product algorithm \cite{dauwels2009,loeliger2007}. In particular, the graphical inference performed by the sum-product algorithm is used to calculate the E-step of the algorithm. (Graphical inference can also be used in the M-step, though we do not use this feature.)

A complete derivation of the EM algorithm for our problem is found in the SI.


\subsection{Simulation}

We test the factor graph EM algorithm via Monte Carlo simulation. 
For our simulations, we discretize time and simulate the receptors using discrete-time Markov chains; this method gives simulated measurements similar to sampled patch clamp traces.

Our figure of merit is {\em probability of error}: for ground-truth state  $s$ and estimate $\hat{s}$, probability of error $P_e$ is defined as
\begin{align}
    P_e = \Pr(s \neq \hat{s}) ,
\end{align}
i.e., the proportion of microstates that are incorrectly detected.

\subsection{Patch clamp measurements}


Single CFTR channels were studied in inside-out patches pulled from {\em Xenopus} oocytes injected with cRNA encoding the wildtype channel, as previously described \cite{InfieldCuiKuangMcCarty2016AmJPhysLung}. 
Briefly, to enable removal of the vitelline membrane, oocytes were placed in a bath solution containing (in mM) 200 monopotassium aspartate, 20 KCl, 1 MgCl\textsubscript{2}, 10 EGTA, and 10 HEPES, pH 7.2 adjusted with KOH.  Gigaohm seals were formed with patch pipettes pulled from borosilicate glass and filled with solution containing (in mM) 150 N-methyl-D-glutamine (NMDG) chloride, 5 MgCl\textsubscript{2}, and 10 TES buffer, pH 7.5.  After excision of the patch, CFTR channels were activated by bath solution containing 150 mM NMDG chloride, 1.1 mM  MgCl\textsubscript{2}, 2 mM Tris-EGTA, 10 mM TES buffer, 1 mM MgATP, and 127 U/ml PKA, pH 7.5.  Currents were recorded at V\textsubscript{M} = -100 mV using an Axopatch 200B amplifier, with filtering at 0.1-1 kHz.

\subsection{Code}

Code implementing the factor graph EM algorithm, both for generating simulations and analyzing experimental patch clamp data, is available on GitHub, and was used to generate all results in this paper: \url{http://github.com/andreweckford/PatchClampFactorGraphEM/}

\section{Results}

We first present results using simulated patch clamp measurements. In these results, the true kinetic state can be obtained from the simulator, so the performance of the algorithm can be analyzed in detail. Subsequently, we apply the algorithm to actual patch clamp traces obtained from a CFTR receptor.

\subsection{Analysis using simulated patch clamp measurements}

We apply our state estimation method to two simulated receptors: cystic fibrosis transmembrane conductance regulator (CFTR) and nicontinic acetylcholine (nAChR). Both receptors have multiple hidden kinetic states in both the open and closed configurations. Beyond the brief descriptions below, detailed descriptions of our simulation environment, and our receptor models, are given in the Methods section and in the SI. 

We use the seven-state CFTR model from \cite{fuller2005block} and the five-state nAChR model from \cite{colquhoun1982}, which are depicted in Figures \ref{fig:CFTR_model} and \ref{fig:ACh_model}, respectively; model parameters are given in Tables \ref{tab:CFTR-parameters} and \ref{tab:ACh-parameters}, respectively. It is important to note that our algorithm works without knowledge of the parameters, and that these parameters are {\em only used to generate simulated patch clamp measurements}. A key difference between the models is that kinetic states in CFTR are ``hard'' to estimate while in nAChR they are ``easy'' (in a way that will be described in detail in the discussion). However, in both cases we show that the factor graph EM algorithm's performance is close to the theoretical optimum.

\begin{table}[h!]
    \centering
    {\scriptsize
    \begin{tabular}{ccccccccc}
    & & &  \rlap{Destination state} & & & & & \\
        & & \conea & \coneb & \ctwo & \oone & \otwo & \cthree & \cfour \\
        \cmidrule{2-9}
        & \conea & &$9.0 \cdot 10^3$ (M s)$^{-1}$ [ATP] & & & & & \\ \cmidrule{2-9}
        & \coneb & 5.0 s$^{-1}$& & 7.7 s$^{-1}$ & & & & \\ \cmidrule{2-9}
        & \ctwo & & 5.8 s$^{-1}$ & & 4.9 s$^{-1}$ & & & \\ \cmidrule{2-9}
        & \oone & & & 10.0 s$^{-1}$ & & 7.1 s$^{-1}$ & & \\ \cmidrule{2-9}
        \rot{\rlap{Origin state}} 
        & \otwo & & & & & & 3.0 s$^{-1}$ & \\ \cmidrule{2-9}
        & \cthree & & & & & 7.0 s$^{-1}$ & & 6.0 s$^{-1}$ \\ \cmidrule{2-9}
        & \cfour & 1.7 s$^{-1}$ & & & & & 12.8 s$^{-1}$ & \\ \cmidrule{2-9}
    \end{tabular}
    }
    \caption{Parameters for the CFTR channel, corresponding to the model in Figure \ref{fig:CFTR_model}. A blank entry indicates that the transition is impossible. [ATP] indicates molar concentration of ATP. 
    }
    \label{tab:CFTR-parameters}
\end{table}

\begin{table}[h!]
    \centering
    {\scriptsize
    \begin{tabular}{ccccccc}
    & & & \rlap{Destination state} & & & \\
        & & O1 & O2 & C3 & C4 & C5 \\ \cmidrule{2-7}
        & O1 & & $5 \cdot 10^8$ (M s)$^{-1}$ [ACh] & & $1 \cdot 10^3$ s$^{-1}$ & \\ \cmidrule{2-7}
        & O2 & 0.66 s$^{-1}$ & & $5 \cdot 10^2$ s$^{-1}$ & \\ \cmidrule{2-7}
        & C3 & & $1.5 \cdot 10^4$ s$^{-1}$ & & $4 \cdot 10^3$ s$^{-1}$ & \\ \cmidrule{2-7}
        \rot{\rlap{Origin state}}
        & C4 & 15 s$^{-1}$ & & $5 \cdot 10^8$ (M s)$^{-1}$ [ACh] & & $2 \cdot 10^3$ s$^{-1}$ \\ \cmidrule{2-7}
        & C5 & & & & $1 \cdot 10^8$ (M s)$^{-1}$ [ACh] & \\ \cmidrule{2-7}
    \end{tabular}
    }
    \caption{Parameters for the nAChR receptor, corresponding to the model in Figure \ref{fig:ACh_model}. A blank entry indicates that the transition is impossible. Parameters are adapted from \cite{colquhoun1982}. [ACh] indicates molar concentration of ACh.}
    \label{tab:ACh-parameters}
\end{table}

In many of these results, we compare a factor graph EM algorithm, \pjt{which} iteratively performs parameter estimation and inference, to factor-graph-based inference, where the model parameters are known in advance by the algorithm. For factor graphs of the type that we consider, the latter method is known to implement maximum {\em a posteriori} probability detection, which delivers the minimum probability of error, on average, of any possible algorithm \cite{kschischang2001}.

\subsubsection{Visualizing the iterative algorithm}

\begin{figure}[h!]
    \centering
    \includegraphics[width=\textwidth]{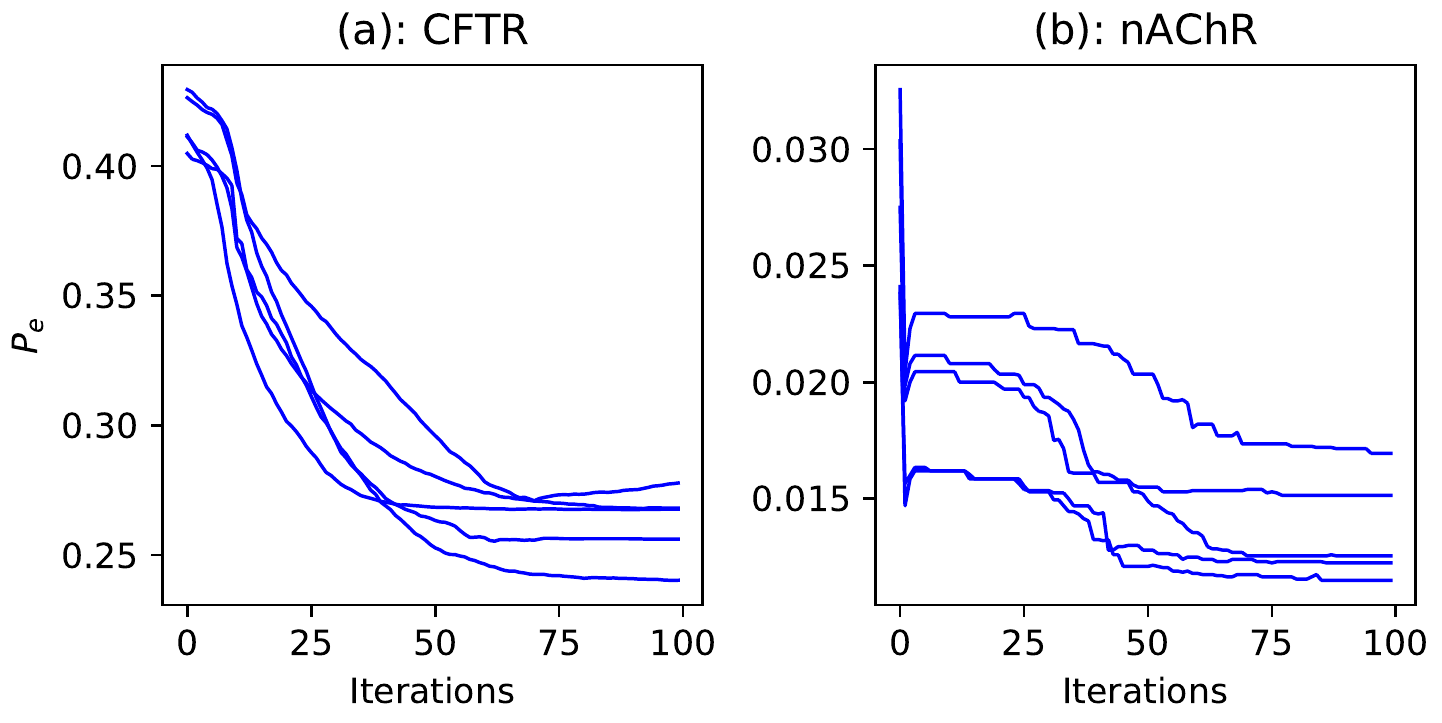}
    \caption{Visualization of the progress of the factor graph EM algorithm, showing the evolution of $P_e$ as a function of iteration number over five trials in CFTR (subfigure (a)), and nAChR (subfigure (b)). Each plot was generated from simulated patch clamp measurements. CFTR: [ATP] = $1.1 \cdot 10^{-3}$ M, sampling rate = 100 Hz; nAChR: [ACh] = $1.0 \cdot 10^{-5}$ M, sampling rate = 20 Hz; 20000 samples for each simulation; rate parameters from Tables \ref{tab:CFTR-parameters} and \ref{tab:ACh-parameters}.}
    \label{fig:EM-visualization}
\end{figure}

The factor graph EM algorithm is iterative, alternating between inference (called the E-step) and estimation of the parameters (called the M-step), with each step updating the other. One full iteration includes both an E-step and an M-step. In Figure \ref{fig:EM-visualization}, we show the progress of the algorithm through 100 iterations, with five simulation runs for each receptor. The probability of error significantly decreases with the number of iterations, although it does not show monotonic behaviour. This may seem surprising since the parameter estimates produced in the M-step are known to be increasing in likelihood with each iteration \cite{dempster1977}. However, the probability of error is calculated with respect to inference performed in the E-step, which has no such guarantee.

From the figure it is apparent that the estimates show near-convergence after 100 iterations, though there are still some small changes in $P_e$. We chose 100 iterations for easy visualization, and not because the algorithm always converges in this timeframe: we have noticed rare cases where the algorithm might still make significant changes after hundreds of iterations, particularly for models with large numbers of parameters like CFTR. For this reason, the rest of our results use larger numbers of iterations.

\subsubsection{Effect of changing agonist concentration}

\begin{figure}[h!]
    \centering
    \includegraphics[width=\textwidth]{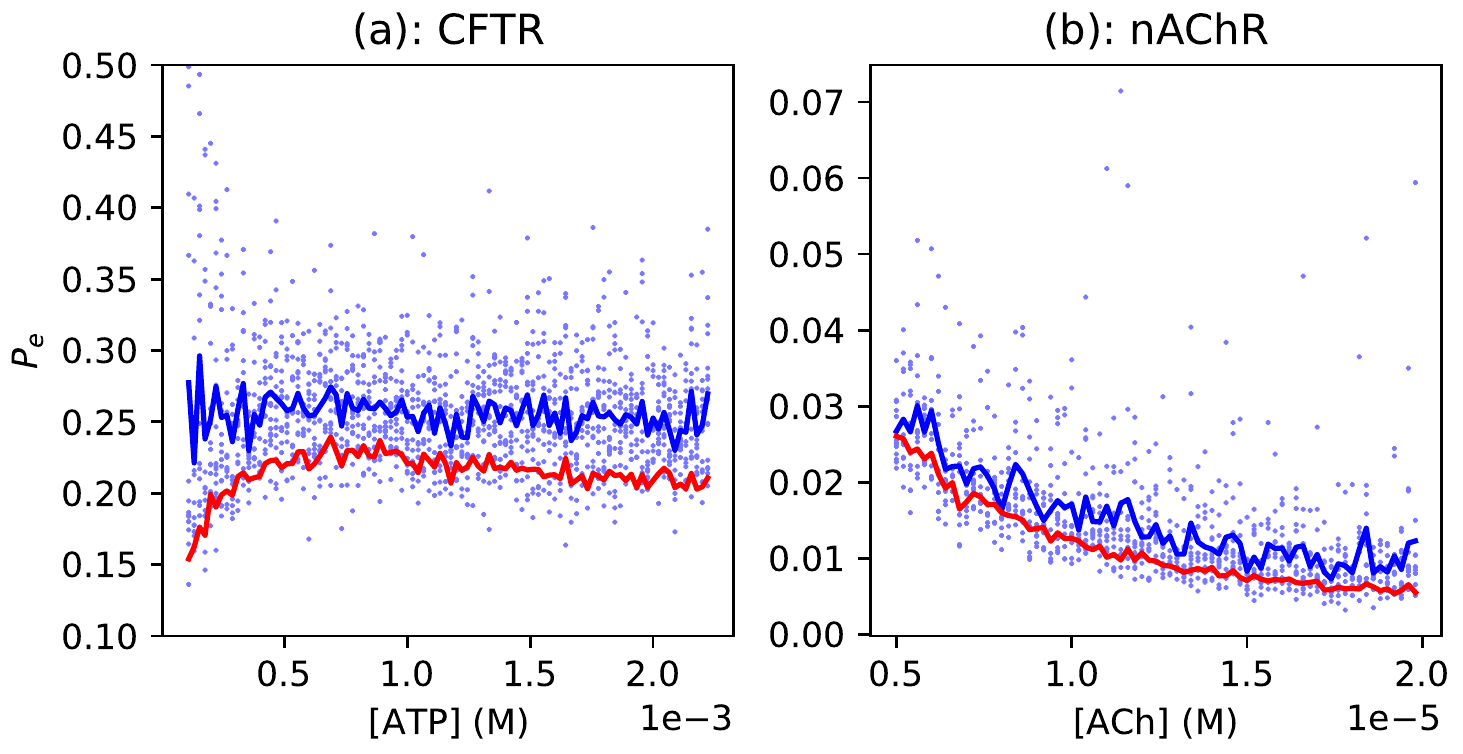}
    \caption{Effect of algorithm performance on changing agonist concentration in CFTR (subfigure (a)) and nAChR (subfigure (b)). {\em Light blue dots:} $P_e$ for the factor graph EM algorithm in each simulation. {\em Blue line:} Average $P_e$ of the factor graph EM algorithm. {\em Red line:} Average $P_e$ of the MAP detector given the true parameter values. CFTR: sampling rate = 100 Hz; nAChR: sampling rate = 20 Hz; 20000 samples and 400 EM iterations for each simulation; rate parameters from Tables \ref{tab:CFTR-parameters} and \ref{tab:ACh-parameters}. In subfigure (a), two light blue points are cut off at upper left.}
    \label{fig:EM-concentration}
\end{figure}

In both models, most of the model parameters are fixed, but a few are sensitive to the concentration of an agonist (see Tables \ref{tab:CFTR-parameters} and \ref{tab:ACh-parameters}): CFTR is sensitive to adenosine triphosphate (ATP), while the nAChR receptor is sensitive to the concentration of acetylcholine (ACh). 
In Figure \ref{fig:EM-concentration}, we show the performance of our algorithm when the concentration of the respective agonist is changed. 

We see that there is an effect on the absolute accuracy of estimation as a function of concentration. 
\pjt{As the concentration of ATP or ACh changes, so does the contribution of distinct microstate transitions to the stationary variance of the recorded current \cite{SchmidtGalanThomas2018PLoSCB}, an example of the stochastic shielding phenomenon \cite{SchmandtGalan2012PRL,SchmidtThomas2014JMatBio}. 
For example, at [ACh]$\sim20\mu$M, the variance in the observable $y$ is dominated by noise arising from the $\otwo\leftrightharpoons\cthree$, while at [ACh]$\sim 5\mu$M, the transitions $\otwo\leftrightharpoons\cthree$, $\cthree\leftrightharpoons\cfour$, and $\cfour\leftrightharpoons\cfive$ make roughly equal contributions to the variance (cf.~\cite{SchmidtGalanThomas2018PLoSCB}, Fig.~2C), which increases the difficulty of unambiguously inferring the microstate given the observed state.}
Nevertheless, the relative accuracy of the EM algorithm, compared with the estimator given the true parameter values, remains relatively constant.

\subsubsection{Discarding low-confidence results}

\begin{figure}[h!]
    \centering
    \includegraphics[width=\textwidth]{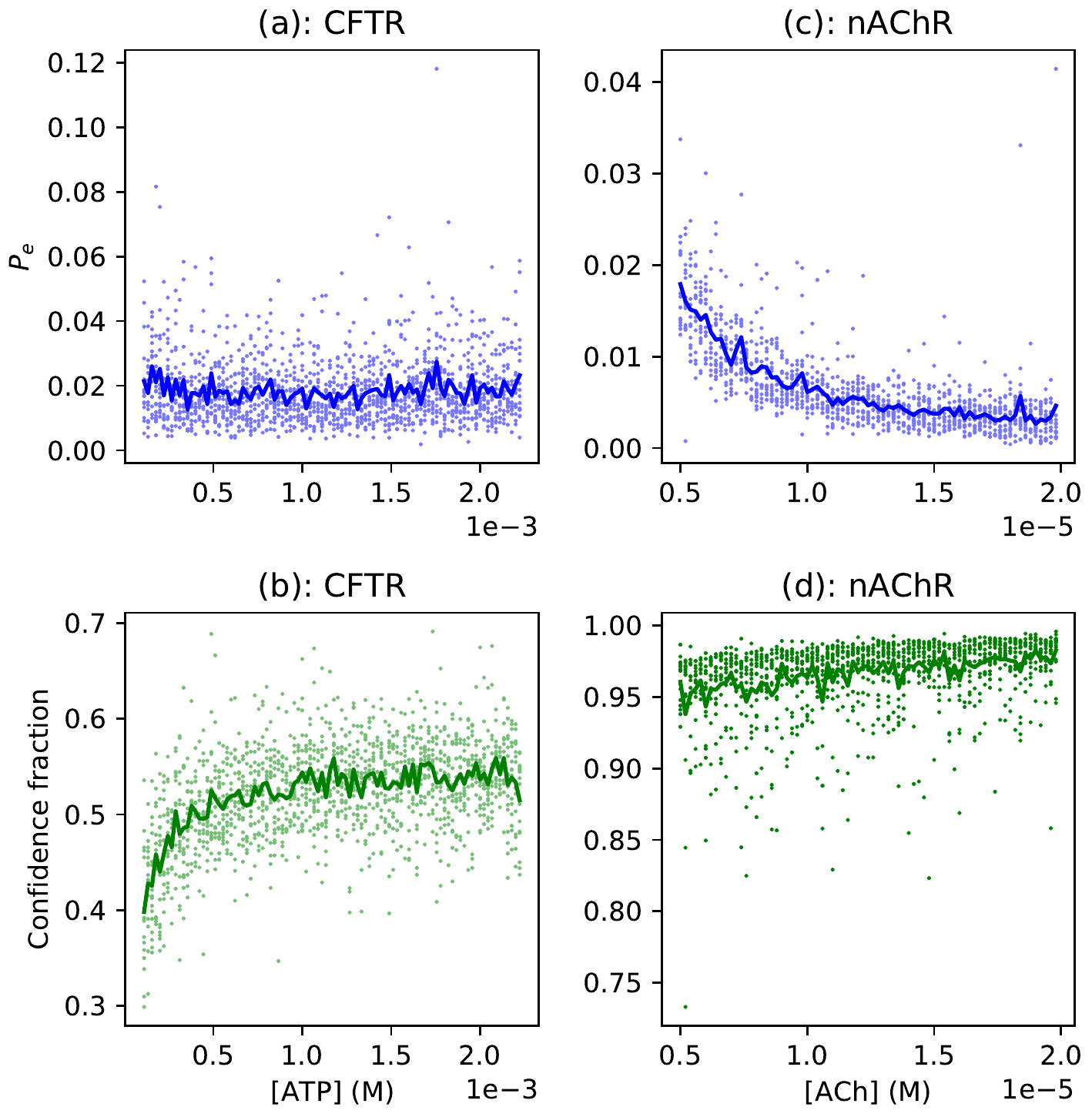}
    \caption{Accuracy of algorithm with low-confidence estimates discarded in CFTR (subfigures (a), (b)) and nAChR (subfigures (c), (d)). In both figures, individually estimated states with a confidence of $c < 0.8$ are discarded. {\bfseries Top row (subfigures (a), (c)):} {\em Light blue dots:} $P_e$ for all estimates with $c \geq 0.8$ in each run. {\em Blue line:} Average $P_e$ of the algorithm for all estimates with $c \geq 0.8$. {\bfseries Bottom row (subfigures (b), (d)):} {\em Light green dots:} Fraction of samples with $c \geq 0.8$ in each run. {\em Green line:} Average fraction of samples with $c \geq 0.8$. CFTR: sampling rate = 100 Hz; nAChR: sampling rate = 20 Hz; 20000 samples and 400 EM iterations for each simulation; rate parameters from Tables \ref{tab:CFTR-parameters} and \ref{tab:ACh-parameters}.}
    \label{fig:EM-confidence}
\end{figure}

The E-step of the factor graph EM algorithm obtains the {\em a posteriori} probability distribution over the hidden kinetic states, subject to: (1) conditioning on all observed patch clamp measurements, and (2) setting the parameter estimates from the M-step equal to the true parameter values. (This is described in more detail in the Methods section.) The factor graph EM algorithm does not obtain the true {\em a posteriori} probability, as that would require knowledge of the system parameters. However, this value can be used as a score of the algorithm's confidence in an estimate, so that estimates with low confidence can be discarded.

In Figure \ref{fig:EM-confidence}, we show the performance of the algorithm when results with confidence score of $c < 0.8$ are excluded. Note that if $c$ was the true {\em a posteriori} probability, then $c = 0.8$ would correspond to $P_e = 1-0.8 = 0.2$. Comparing Figure \ref{fig:EM-confidence} to Figure \ref{fig:EM-concentration}, with CFTR the $P_e$ is significantly greater than the chosen $c$, so if $c$ is a good approximation for $P_e$, relatively few results would have high confidence. Indeed, this behaviour is apparent from the CFTR figure: for estimates with $c \geq 0.8$, we have $P_e \simeq 0.2$ on average, but only a third (or less) of the time series achieves $c \geq  0.8$. 
Meanwhile, with nAChR, there is little improvement in performance since $P_e$ was already quite low; however, very few estimates are rejected with $c < 0.8$ (the number of rejected estimates is so low that we do not plot it in this figure).




\subsubsection{Noisy measurements}
\label{sec:noisy-results}

\begin{figure}[h!]
    \centering
    \includegraphics[width=\textwidth]{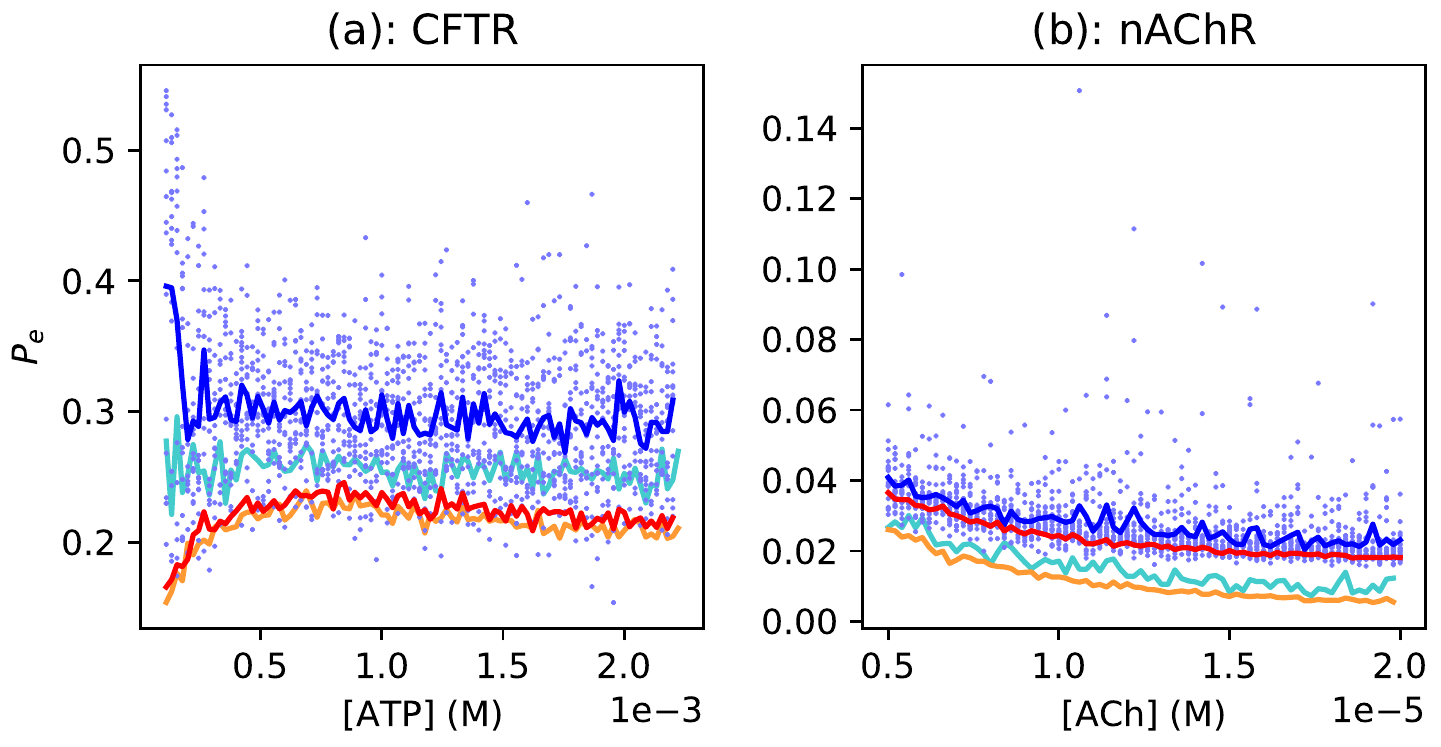}
    \caption{Accuracy of the algorithm for changing input concentration with noisy patch clamp measurements for CFTR (subfigure (a)) and nAChR (subfigure (b)); setup is similar to Figure \ref{fig:EM-concentration} but Gaussian noise with $\sigma^2 = 0.1$ is added. {\em Light blue dots:} $P_e$ for the factor graph EM algorithm in each simulation. {\em Blue line:} Average $P_e$ of the factor graph EM algorithm. {\em Red line:} Average $P_e$ of the MAP detector given the true parameter values.  {\em Orange and cyan lines}: Identical to the red and blue lines (respectively) from Figure \ref{fig:EM-concentration}, for comparison. CFTR: sampling rate = 100 Hz; nAChR: sampling rate = 20 Hz; 20000 samples and 400 EM iterations for each simulation; rate parameters from Tables \ref{tab:CFTR-parameters} and \ref{tab:ACh-parameters}.}
    \label{fig:EM-noisy}
\end{figure}

In the previous results, the algorithm was provided noiseless knowledge of the state of the ion channel, i.e., the algorithm could tell perfectly if the ion channel was open or closed. While this is a useful assumption for determining the best possible performance of the algorithm, and for illustrating our approach, realistic patch clamp measurements are noisy.

We use an additive Gaussian noise model to represent the patch clamp measurement. That is, the patch clamp observes $Y$, where
\begin{align}
    Y &= I_{s_k} + N,
\end{align}
in which $I_{s_k}$ (in picoamps) is the current corresponding to kinetic state $s_k$, and $N$ (in picoamps) is additive Gaussian noise with variance $\sigma^2$. The EM algorithm then estimates both the transition probability matrix $P$ and the noise variance $\sigma^2$, a detailed algorithm for which is provided in the SI. Results are shown in Figure \ref{fig:EM-noisy}, in which $\sigma^2 = 0.1$. As expected, the noise degrades the accuracy of the estimates, but the EM algorithm still performs well compared with a sum-product algorithm with known parameters. 

\subsection{Application to patch clamp measurements}

\begin{figure}[h!]
    \centering
    \includegraphics[width=\textwidth]{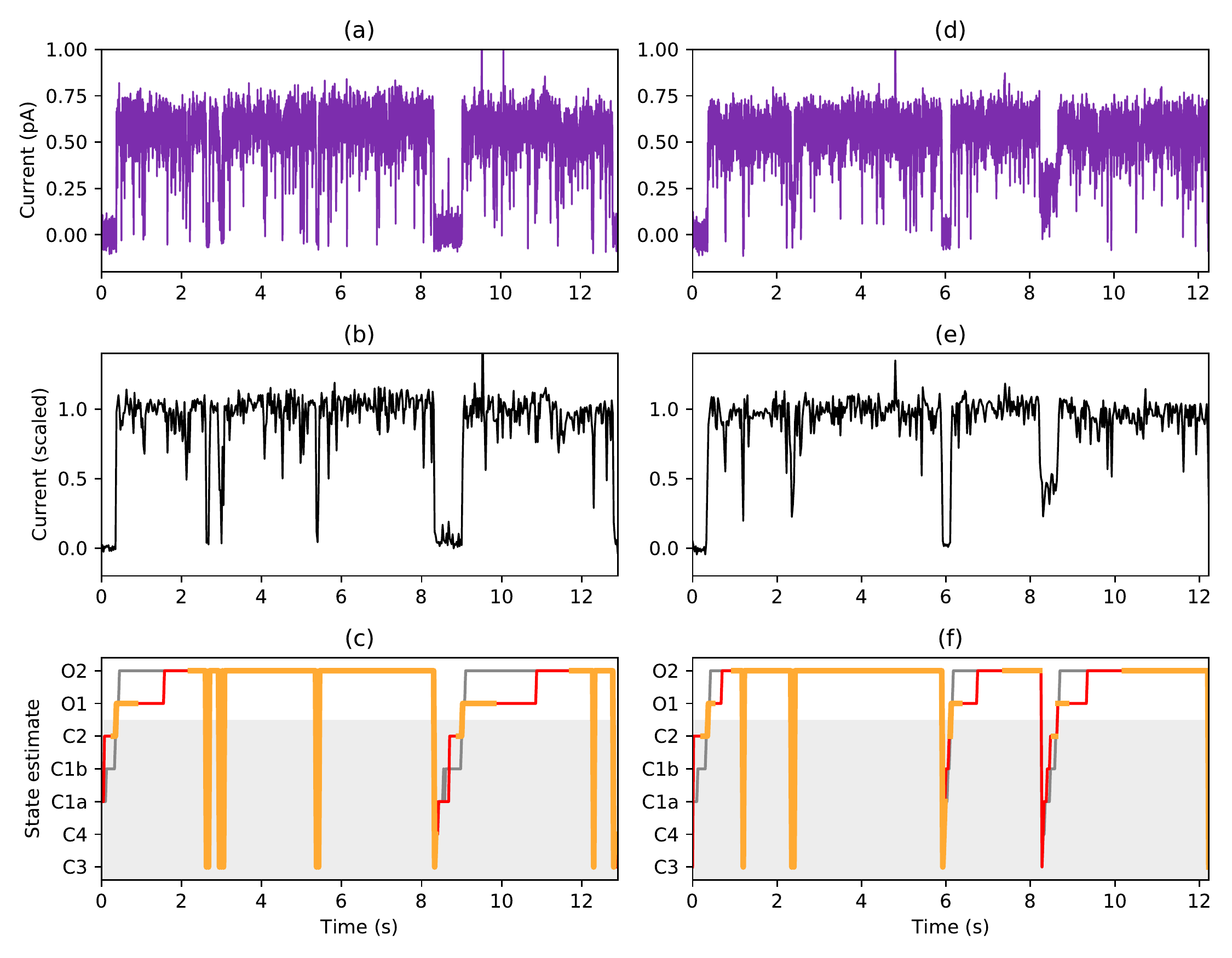}
    \caption{CFTR patch clamp measurements (raw and preprocessed) along with the corresponding hidden state estimates for two different experiments, one in each column.
    {\bfseries Top row (subfigures (a), (d)):} {\em Purple line:} measured patch clamp current.
    {\bfseries Middle row (subfigures (b), (e)):} {\em Black line:} patch clamp current signal after filtering and preprocessing; this signal is provided to the factor graph EM algorithm. {\bfseries Bottom row (subfigures (c), (f)):} {\em Grey line:} estimated kinetic state after 10 EM iterations; {\em orange/red line:} estimated kinetic state after 100 EM iterations, where orange points exceed confidence threshold of 0.8, and red points are below confidence threshold. In the bottom plots, open channel states have a white background, while closed channel states have a light grey background.}
    \label{fig:patchclamp}
\end{figure}

In Figure \ref{fig:patchclamp}, we show the application of our algorithm to CFTR patch clamp measurements. 
We show two examples, corresponding to two different experiments. 
For techniques used to obtain these measurements, see the Methods section.

Prior to applying the patch clamp signal to the factor graph EM algorithm, we perform a preprocessing step. This step consists of block averaging (i.e., taking the sample average over non-overlapping blocks of 50 samples) and scaling (multiplying by a scalar, here 1.75, so that most signal features are in the range from 0 to 1). This step is performed to reduce noise and improve the performance of the algorithm.

In the bottom plots, we give the patch clamp measurement (light blue line), estimate after 10 EM iterations (grey line), and the estimate after 100 EM iterations (red line). Estimates judged with high confidence (as described above) are highlighted as orange lines. Results from different numbers of iterations are presented to illustrate the progress of the algorithm, while the result from the highest number of iterations would normally be used. From the raw data and preprocessed traces (purple and black lines, respectively), the periods in which the ion channel admits higher current correspond well to the times when the EM algorithm estimates an open state. Moreover, the EM algorithm gives useful estimates of hidden states.

It should be emphasized that the hidden state estimate is the most likely estimate at each time, and might not have the features of a valid trajectory. For example, the EM algorithm rarely decides that CFTR is in state \cfour; however, the mechanics of the receptor mean that most valid trajectories should pass through \cfour. This may be because \cfour{} is a briefly-occupied state, \pjt{and because observations into and out of $\cfour$ are not directly observable \cite{eckford2018channel}.  Thus} 
the EM algorithm does not have the opportunity to gather evidence that the receptor is occupying this state.



\section{Discussion}

The model-based factor graph EM algorithm approach is highly flexible and extensible. In terms of receptors, while we presented two examples in this paper (CFTR and nAChR), the approach could in principle be used on any receptor with an ion channel, as long as a kinetic model is available. Notably, machine learning approaches to approximate inference can incorporate the EM algorithm \cite{goodfellow2016deep}. In terms of techniques, the inference-based approach can be easily extended for use in combination with other instruments or techniques as long as a stochastic model exists for them. Moreover, the factor graph EM algorithm has significant advantages. It is relatively easy to derive and implement, and no training phase is required. Further, as indicated in our results, the performance in terms of $P_e$ is close to the theoretical optimum.

The algorithm presented in this paper is applied retrospectively to a complete patch clamp trace collected in the laboratory. 
As such, if particular kinetic microstates are of interest in an experiment, our algorithm can be used to identify when those microstates occur, measuring the effect of any interventions. 
This \pjt{feature} gives the designer of an experiment a novel and fine-grained algorithmic tool to discover changes to the behaviour of receptor proteins. For example, this method could be used to determine the fine-grained, microstate-by-microstate effects of particular agonists or mutations.

Looking towards future work, reliable estimation of the hidden kinetic state of a receptor could lead to new experimental approaches by targeting those individual states in real time. With adjustments, especially incorporating recursive or ``on-line'' variants of the EM algorithm \cite{cappe2009line}, our method could be adapted to provide immediate estimates of the hidden state while an experiment is ongoing. An intervention specific to the microstate, for example using targeted electromagnetic radiation \cite{turton2014terahertz,elayan2021information}, could then be used to influence the hidden state, either forcing it to exit or remain in the targeted state. In principle, this method could also be used to design new therapeutic interventions.

\section{Conclusion}

We have demonstrated that the factor graph EM algorithm is an accurate method for estimating the hidden kinetic states from patch clamp current measurements\pjt{, given an underlying kinetic model with unknown parameters}. 
The methods demonstrated in this paper may be applied widely, to virtually any ion channel. Our work opens up the possibility of novel experimental designs that can exploit fine-grained kinetic state estimation. There is also a rich potential for future theoretical work, such as adapting the algorithm to incorporate other observations alongside the patch clamp, and deriving an algorithm that can handle multiple simultaneous open channels.








\section*{Acknowledgments}

ASM and AWE were funded by DARPA via the RadioBio program under grant number HR001117C0125. 
PJT was supported by NSF via grant number DMS-2052109 and by Oberlin College.
WDH, NAM, and RSW were funded in part by DARPA via the RadioBio program under grant number HR001117C0124.


%
%
%

\renewcommand{\baselinestretch}{1}
\normalsize
\bibliographystyle{unsrt}
\bibliography{refs}

\clearpage

\renewcommand{\baselinestretch}{1.5}
\normalsize

\renewcommand{\theequation}{S\arabic{equation}}
\setcounter{equation}{0}
\renewcommand{\thefigure}{S\arabic{figure}}
\setcounter{figure}{0}
\renewcommand{\thesection}{S\arabic{section}}
\setcounter{section}{0}

\section*{Supplemental Information}



\section{Master equations}

\subsection{CFTR}

Consider the master equation from equation (\ref{eqn:master}) in the main paper. Here $Y$ gives the occupancy probabilities for each microstate. In the case of the CFTR receptor, 
\begin{align}
    \label{eqn:CFTR-Y}
    Y = \Big[ P(\conea),P(\coneb),P(\ctwo),P(\oone),P(\otwo),P(\cthree),P(\cfour)\Big] .
\end{align}
Using (\ref{eqn:CFTR-Y}) and equation (\ref{eqn:R-CFTR}) in the main paper, the continuous-time master equation is fully expanded by 
\begin{align}
    \label{eqn:CFTR_MEQ_C1a}
    \frac{dP(\conea)}{dt}&=-k_{\conea\rightarrow{}\coneb}P(\conea)+k_{\cfour\rightarrow{}\conea}P(\cfour)+k_{\coneb\rightarrow{}\conea}P(\coneb)\\
    \label{eqn:CFTR_MEQ_C1b}
    \frac{dP(\coneb)}{dt}&=-(k_{\coneb\rightarrow{}\ctwo}+k_{\coneb\rightarrow{}\conea})P(\coneb)+k_{\conea\rightarrow{}\coneb}P(\conea)+k_{\ctwo\rightarrow{}\coneb}P(\ctwo)\\
    \label{eqn:CFTR_MEQ_C2}
    \frac{dP(\ctwo)}{dt}&=-(k_{\ctwo\rightarrow{}\oone}+k_{\ctwo\rightarrow{}\coneb})P(\ctwo)+k_{\coneb\rightarrow{}\ctwo}P(\conea)+k_{\oone\rightarrow{}\ctwo}P(\oone)\\
    \label{eqn:CFTR_MEQ_O1}
    \frac{dP(\oone)}{dt}&=-(k_{\oone\rightarrow{}\otwo}+k_{\oone\rightarrow{}\ctwo})P(\oone)+k_{\ctwo\rightarrow{}\oone}P(\ctwo)\\
    \label{eqn:CFTR_MEQ_O2}
    \frac{dP(\otwo)}{dt}&=-k_{\otwo\rightarrow{}\cthree}P(\otwo)+k_{\oone\rightarrow{}\otwo}P(\oone)+k_{\cthree\rightarrow{}\otwo}P(\cthree)\\
    \label{eqn:CFTR_MEQ_C3}
    \frac{dP(\cthree)}{dt}&=-(k_{\cthree\rightarrow{}\cfour}+k_{\cthree\rightarrow{}\otwo})P(\cthree)+k_{\otwo\rightarrow{}\cthree}P(\otwo)+k_{\cfour\rightarrow{}\cthree}P(\cfour)\\
    \label{eqn:CFTR_MEQ_C4}
    \frac{dP(\cfour)}{dt}&=-(k_{\cfour\rightarrow{}\conea}+k_{\cfour\rightarrow{}\cthree})P(\cfour)+k_{\cthree\rightarrow{}\cfour}P(\cthree) .
\end{align}
Example values of each rate are given in Table \ref{tab:CFTR-parameters} in the main paper. We suppress the dependence of $k_{\conea\rightarrow\coneb}$ on ATP concentration in our notation for the sake of compactness, but as noted in Table \ref{tab:CFTR-parameters} the effects of ATP concentration are fully considered in the model.

\subsection{nAChR}

Performing a similar expansion to the above, now $Y$ gives the occupancy probabilities for nAChR microstates, i.e., 
\begin{align}
    \label{eqn:ACh-Y}
    Y = \Big[ P(\oone),P(\otwo),P(\cthree),P(\cfour),P(\cfive)\Big] .
\end{align}
The continuous-time master equation describing the nAChR cycle is
\begin{align}
  \label{eqn:ACh_MEQ_O1}
  \frac{dP(\oone)}{dt}&=-(k_{\oone\rightarrow{}\otwo}+k_{\oone\rightarrow{}\cfour})P(\oone)+k_{\cfour\rightarrow\oone}P(\cfour)+k_{\otwo\rightarrow\oone}P(\otwo)\\
  \label{eqn:ACh_MEQ_O2}
  \frac{dP(\otwo)}{dt}&=-(k_{\otwo\rightarrow{}\cthree}+k_{\otwo\rightarrow{}\oone})P(\otwo)+k_{\oone\rightarrow\otwo}P(\oone)+k_{\cthree\rightarrow\otwo}P(\cthree)\\
  \label{eqn:ACh_MEQ_C3}
  \frac{dP(\cthree)}{dt}&=-(k_{\cthree\rightarrow{}\cfour}+k_{\cthree\rightarrow{}\otwo})P(\cthree)+k_{\otwo\rightarrow\cthree}P(\otwo)+k_{\cfour\rightarrow\cthree}P(\cfour)\\
  \label{eqn:ACh_MEQ_C4}
  \frac{dP(\cfour)}{dt}&=-(k_{\cfour\rightarrow{}\cfive}+k_{\cfour\rightarrow{}\oone}+k_{\cfour\rightarrow{}\cthree})P(\cfour)\\
  \nonumber
  &\:\:\:\:+k_{\cthree\rightarrow\cfour}P(\cthree)+k_{\oone\rightarrow\cfour}P(\oone)+k_{\cfive\rightarrow\cfour}P(\cfive)\\
  \label{eqn:ACh_MEQ_C5}
  \frac{dP(\cfive)}{dt}&=-k_{\cfive\rightarrow{}\cfour}P(\cfive)+k_{\cfour\rightarrow{}\cfive}P(\cfive) .
\end{align}
Example rates are described in Table \ref{tab:ACh-parameters} in the main paper. As with the dependence of our CFTR model on ATP concentration, we suppress the dependence of $k_{\oone\rightarrow\otwo}$, $k_{\cfive\rightarrow\cfour}$, and $k_{\cfour\rightarrow\cthree}$ on acetylecholine concentration in our notation.

\section{Factor graphs and the sum-product algorithm}
\label{sec:SumProductDerivation}

\subsection{General considerations}

Factor graphs and the sum-product algorithm are general methods for probabilistic inference, and can in principle be applied to any inference problem using a given stochastic model. Here, for brevity, we focus on the special case of inference in our CFTR model; for more details and generalizations, the reader is encouraged to consult \cite{kschischang2001}.

\begin{figure}
    \centering
    \includegraphics[width=\textwidth]{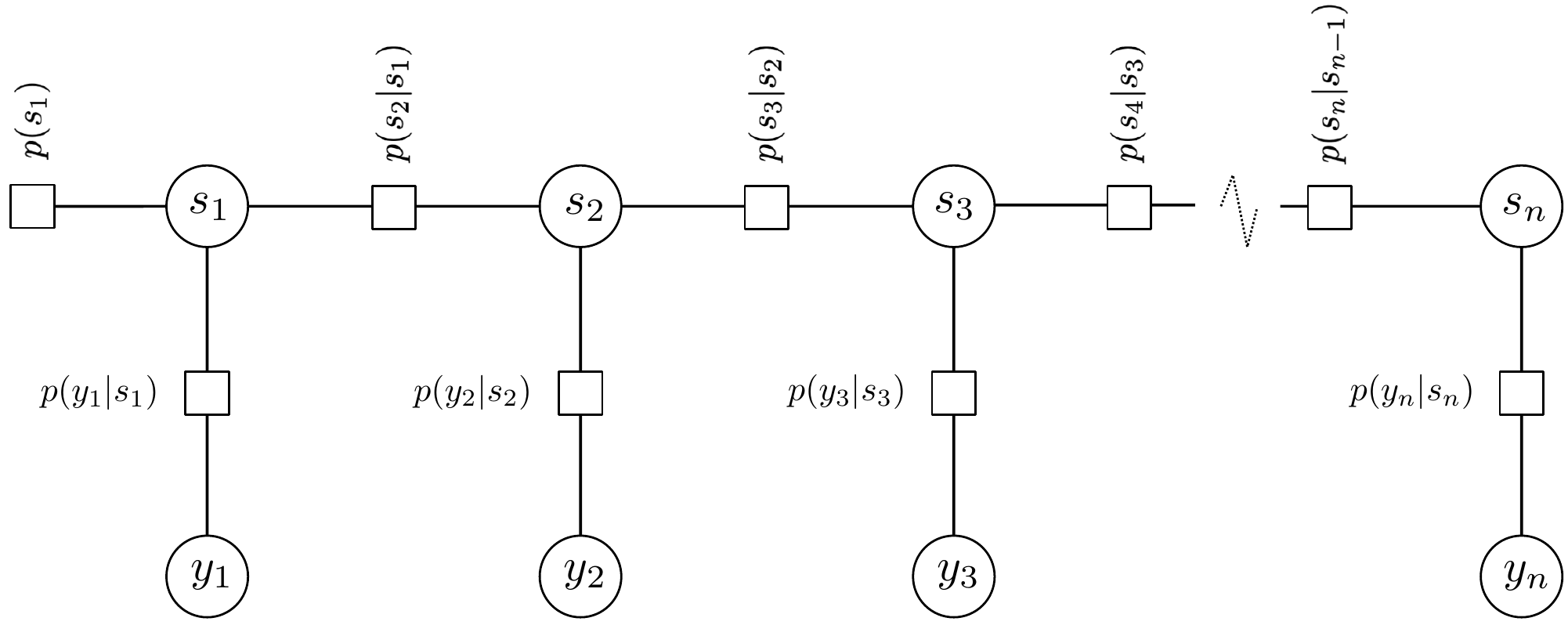}
    \caption{Factor graph for the patch clamp observations.}
    \label{fig:FactorGraphBare}
\end{figure}

Consider first the case of a single channel. The probability mass function of $s$ can be written
\begin{align}
    p(s) &= \prod_{k=1}^n p(s_k \given s_{k-1}) ,
\end{align}
where $s_0$ is null, i.e., $p(s_1 \given s_0) = p(s_1)$. We can include $y$ in the probabilistic model: since $y_k$ is deterministic given $s_k$, the probability of $y_k$ given $s_k$ is the Kronecker delta function between $y_k$ and $m(s_k)$:
\begin{align}
    p(y_k \given s_k) &= 
    \left\{ 
        \begin{array}{cl} 
            1, & y_k = m(s_k) \\ 
            0, & y_k \neq m(s_k) 
        \end{array}
    \right.
\end{align}
Finally, we have the joint probability mass function
\begin{align}
    \label{eqn:factorization}
    p(y,s) &= \prod_{k=1}^n p(y_k \given s_k) p(s_k \given s_{k-1}) .
\end{align}
The stochastic model $p(y,s)$ can be represented on a {\em factor graph}, where nodes representing variables $s_1,\ldots,s_n$ and $y_1,\ldots,y_n$ are connected to nodes representing factors in (\ref{eqn:factorization}), and an edge is drawn from variable to factor if the factor is a function of the variable. The factor graph for (\ref{eqn:factorization}) is depicted in Figure \ref{fig:FactorGraphBare}.

\begin{figure}
    \centering
    \includegraphics[width=\textwidth]{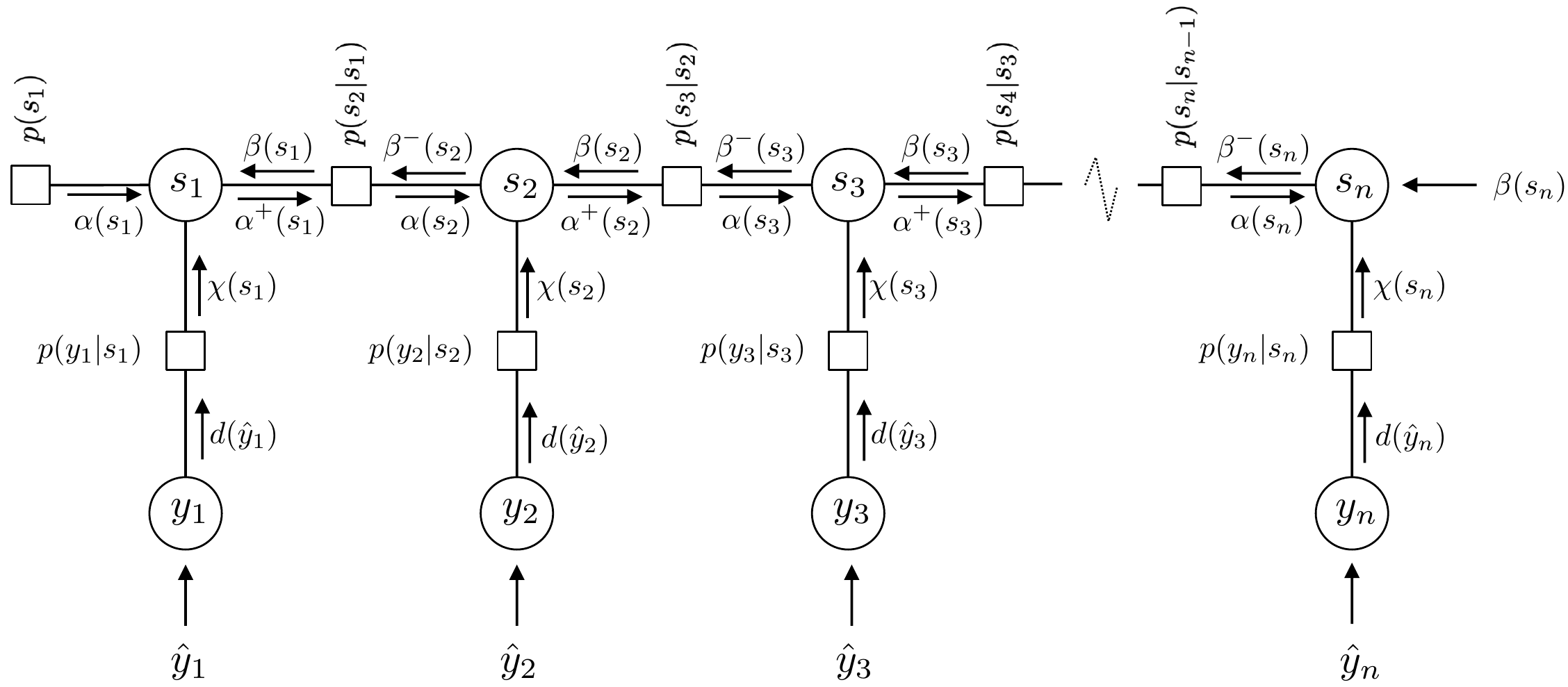}
    \caption{Factor graph for the patch clamp observations with sum-product messages indicated on each edge (cf. Figure \ref{fig:FactorGraphBare}).}
    \label{fig:FactorGraph}
\end{figure}


We are interested in the {\em a posteriori} probability $p(s_k \given y)$, i.e., the distribution over $s_k$ knowing the observable state for all time. This can be calculated using the {\em sum-product algorithm}, which is a message-passing algorithm over the factor graph. 

The flow of messages within the factor graph is depicted in Figure \ref{fig:FactorGraph}. {\em Generally speaking, messages on the outgoing edges from a node are calculated using all incoming messages to that node, {\bfseries except} the incoming message along the same outgoing edge.}

In terms of the variable nodes $y_k$, $s_k$:
\begin{itemize}
    \item At $y_k$, let $\hat{y}_k$ represent the observed value of $y_k$. Then the outgoing message from $y_k$ is the Kronecker delta function $\delta(y_k,\hat{y}_k)$, as the probability is 1 that $y_k = \hat{y}_k$. Since $y_k,\hat{y}_k \in \{0,1\}$, this can be represented as a vector $d(\hat{y}_k) = [\delta(0,\hat{y}_k),\delta(1,\hat{y}_k)]$, i.e.,:
    \begin{align}
        d(\hat{y}_k) &= 
        \left\{
            \begin{array}{cl}
                \ [1, 0]^T, & \hat{y}_k = 0 \\
                \ [0, 1]^T, & \hat{y}_k = 1
            \end{array}
        \right. 
    \end{align}
    
    \item At $s_k$, the forward message $\alpha_v^+(s_k)$ is the component-wise product of the incoming forward message $\alpha_v(s_k)$ and the channel message $\chi(s_k)$\pjt{, defined below}; the backward message $\beta_v^-(s_k)$ is the component-wise product of the incoming backward message $\beta_v(s_k)$ and the channel message $\chi(s_k)$.
    
    It will be convenient to express the forward messages $\alpha$ as row vectors, and the backward messages $\beta$ as column vectors.
    
    Representing these message calculations as matrix multiplications,
    \begin{align}
        \label{eqn:varfwd}
        \alpha_v^+(s_k) &= \alpha_v(s_k) \text{diag}\Big(\chi(s_k)\Big) \\
        \label{eqn:varback}
        \beta_v^-(s_k) &= \text{diag} \Big( \chi(s_k) \Big) \beta_v(s_k)
    \end{align}
\end{itemize}

In terms of the factor nodes $p(y_k \given s_k)$, $p(s_k \given s_{k-1})$, each factor node ``contains'' the factor, and this factor is incorporated into the message calculation. As the factors are conditional probability functions in two variables, it will be convenient to represent the factors as matrices, with the conditional variable remaining constant along the rows, and the probabilistic variable remaining constant along the columns; this implies the matrices are row-stochastic. Message calculations are performed as follows:
\begin{itemize}
    \item At $p(y_k \given s_k)$, the factor can be represented by a $|\mathcal{S}|\times|\mathcal{Y}|$ matrix $M(s_k,y_k)$:
    \begin{align}
        M(y_k,s_k) &= \Big[ M_{ij}(s_k,y_k) \Big] 
    \end{align}
    where
    \begin{align}
        M_{ij}(s_k,y_k) &= \Pr(y_k = j \given s_k = i)
    \end{align}
    with slight abuse of the notation: $y_k = j$ means $y_k$ is the $j$th element of $\mathcal{Y}$ from (\ref{eqn:Yset}), and $s_k = i$ means $x_k$ is the $i$th element of $\mathcal{S}$ from (\ref{eqn:Sset}). 
    The channel message $\chi(s_k)$ is given by the matrix product 
    \begin{align}
        \chi(s_k) &= M(s_k,y_k) d(\hat{y}_k) .
    \end{align}
    
    \item At $p(s_k \given s_{k-1})$, we have an outgoing forward message $\alpha(s_k)$ and an outgoing backward message $\beta(s_{k-1})$, with an incoming forward message $\alpha^+(s_{k-1})$ and incoming backward message $\beta^-(s_k)$. The internal factor $p(s_k \given s_{k-1})$ is represented by the transition probability matrix $P$. The message calculation is given by 
    \begin{align}
        \label{eqn:facfwd}
        \alpha(s_k) &= \alpha^+(s_{k-1}) P \\
        \label{eqn:facback}
        \beta(s_{k-1}) &= P \beta^-(s_k)
    \end{align}
\end{itemize}
Combining (\ref{eqn:varfwd}) and (\ref{eqn:facfwd}), and (\ref{eqn:varback}) and (\ref{eqn:facback}), we can write 
\begin{align}
    \label{eqn:fwdfull}
    \alpha(s_k) &= \alpha(s_{k-1})\text{diag}\Big( \chi(s_{k-1}) \Big) P\\
    \label{eqn:backfull}
    \beta(s_k) &= P \text{diag}\Big( \chi(s_{k+1}) \Big) \beta(s_{k+1})
\end{align}

Next we must specify the order in which operations occur, i.e., the message-passing schedule:
\begin{itemize}
    \item The channel messages $\chi(s_k)$ are precalculated. 
    \item For the forward messages, the initial forward message $\alpha(s_1)$ is the steady-state distribution associated with $P$. We then iteratively calculate (\ref{eqn:fwdfull}) for each $k = 2,3,\ldots$. 
    \item For the backward messages, we set the initial backward message $\beta(s_n) = [1,1,\ldots,1]$ (this message is uninformative, as the future gives us no information). We then iteratively calculate (\ref{eqn:backfull}) for each $k=n-1,n-2,\ldots$.  
\end{itemize}

Finally, to calculate the {\em a posteriori} probability $p(s_k \given y)$, we perform the complete message-passing schedule described above. Following the message calculations, we take the product of all messages inbound to variable node $s_k$, and normalize (to force the product to sum to 1). Letting $\odot$ represent componentwise multiplication:
\begin{align}
    \label{eqn:aposteriori}
    p(s_k \given y) &= \frac{\alpha(s_k) \odot \beta(s_k) \odot \chi(s_k)}{\sum_{s_k} \alpha(s_k) \odot \beta(s_k) \odot \chi(s_k)} .
\end{align}
(Normalizations can also be performed at any intermediate stage of the message calculations, and this is often done for numerical stability.)

\section{Factor Graph EM algorithm}

The message-passing algorithm given in Section \ref{sec:SumProductDerivation} requires knowledge of the system parameters, especially the state transition probability matrix $P$. (If the patch clamp observations are corrupted by noise, we must also estimate the noise variance $\sigma^2$.) As discussed in the main paper, the factor graph EM algorithm incorporates sum-product message passing in each iteration. The algorithm used for our main results is derived below.

\subsection{Noise-free patch clamp observations}

With observations $y$, hidden states $s$, matrix $P$ of parameters (i.e.~the state transition probability matrix), and estimate $\bar{P}$ of the parameters, we have
\begin{align}
    \label{eqn:Qfunction}
    Q(P;\bar{P}) &= E_{\bar{P}} \Big[ \log p(y,s;P) \given y \Big]
\end{align}
where the subscript $\bar{P}$ indicates that the expectation is taken while setting $P = \bar{P}$. From (\ref{eqn:factorization}) we can write
\begin{align}
    \nonumber \lefteqn{E_{\bar{P}} \Big[ \log p(y,s;P) \given y \Big]} & \\
    &= E_{\bar{P}} \left[ \log \prod_{k=1}^n p(y_k \given s_k) p(s_k \given s_{k-1}; P) \given y \right]\\
    \label{eqn:em-derivation-1}
    &= \sum_{k=1}^n E_{\bar{P}} \Big[ \log p(y_k \given s_k) \given y \Big]
        + \sum_{k=1}^n E_{\bar{P}} \Big[ \log p(s_k \given s_{k-1};P) \given y \Big] .
\end{align}
The first term in (\ref{eqn:em-derivation-1}) is constant with respect to $P$ and is not important for the rest of the derivation, so we will absorb it into a constant $C$. Now we have
\begin{align}
    \nonumber \lefteqn{E_{\bar{P}} \Big[ \log p(y,s;P) \given y \Big]} & \\
    &= \sum_{k=1}^n E_{\bar{P}} \Big[ \log p(s_k \given s_{k-1};P) \given y \Big] + C \\
    \label{eqn:em-derivation-2}
    &= \sum_{k=1}^n \sum_{s_k,s_{k-1}} p(s_k,s_{k-1} \given y;\bar{P}) \log p(s_k \given s_{k-1};P) + C .
\end{align}
The term $p(s_k,s_{k-1} \given y;\bar{P})$ can be obtained directly from the sum-product algorithm, as the (normalized) product of all messages incident to the factor node $p(s_k \given s_{k-1})$, setting $P = \bar{P}$ throughout the factor graph. That is, forming a $|\mathcal{S}|\times|\mathcal{S}|$ matrix $Q^{(k)} = [Q_{ij}^{(k)}]$, where 
\begin{align}
    \label{eqn:estep}
    Q^{(k)} &= \text{diag}(\alpha^+(s_{k-1})) \bar{P} \text{diag}(\beta^-(s_k)) ,
\end{align}
we have that $Q_{ij}^{(k)} = p(s_k = j, s_{k-1} = i \given y;\bar{P})$, i.e., we can read the values of $p(s_k,s_{k-1} \given y;\bar{P})$ from $Q^{(k)}$. Finally, returning to (\ref{eqn:em-derivation-2}),
\begin{align}
    \nonumber \lefteqn{E_{\bar{P}} \Big[ \log p(y,s;P) \given y \Big]} & \\
    \label{eqn:em-derivation-3}
    &= \sum_{k=1}^n \sum_{s_k,s_{k-1} = (j,i)} Q^{(k)}_{ij} \log P_{ij} + C .
\end{align}
Calculation of (\ref{eqn:em-derivation-3}) constitutes the E-step of the EM algorithm.

The M-step is performed as follows. Recall that $P$ is row-stochastic, so for constant $i$, $P_{ij}$ forms a probability mass function in $j$. With some manipulation (\ref{eqn:em-derivation-3}) can be rewritten
    \begin{align}
        \nonumber \lefteqn{E_{\bar{P}} \Big[ \log p(y,s;P) \given y \Big]} & \\
        &= \sum_{s_{k-1}} \sum_{s_k} \sum_{k=1}^n Q^{(k)}_{ij} \log P_{ij} + C \\
        &= \sum_{s_{k-1}} Z_i \sum_{s_k} \frac{Q_{ij}}{Z_i} \log P_{ij} ,
    \end{align}
    where
    \begin{align}
        Q_{ij} &= \sum_{k=1}^n Q_{ij}^{(k)} \\
        Z_i &= \sum_{s_k} Q_{ij}
    \end{align}
    Each inner sum of the form $\sum_{s_k} \frac{Q_{ij}}{Z_i} \log P_{ij}$ is maximized by setting
    \begin{align}
        \label{eqn:em-derivation-4}
        P_{ij} &= \frac{Q_{ij}}{Z_i} .
    \end{align}
    Thus, the M-step is accomplished by forming the matrix $P = [P_{ij}]$, with $P_{ij}$ given by (\ref{eqn:em-derivation-4}).
    
\subsection{Patch clamp observations in Gaussian noise with unknown variance}

For the results in Figure \ref{fig:EM-noisy} in the main paper, we assume that the patch clamp observations could be modelled as 
\begin{align}
    \label{eqn:awgn}
    Y &= I + N,
\end{align}
where $I$ is the patch clamp current, and $N$ is an additive white Gaussian noise with zero mean and variance $\sigma^2$. Further we assume that $\sigma^2$ is unknown to the algorithm, and must be estimated in addition to $P$.

Here we derive the EM algorithm that joinly estimates $P$ and $\sigma^2$.
Modifying (\ref{eqn:em-derivation-1}), we can write
\begin{align}
    \nonumber
    \lefteqn{E_{\bar{P},\bar{\sigma}^2} \Big[ \log p(y,s; P,\sigma^2) \Big]} & \\
    \label{eqn:em-noise-derivation-1}
    &= \sum_{k=1}^n E_{\bar{P},\bar{\sigma}^2} \Big[ \log p(y_k \given s_k; \sigma^2) \given y \Big]
        + \sum_{k=1}^n E_{\bar{P},\bar{\sigma}^2} \Big[ \log p(s_k \given s_{k-1};P) \given y \Big] .
\end{align}
In (\ref{eqn:em-noise-derivation-1}), note that only the first term is a function of $\sigma^2$, and only the second term is a function of $P$. Thus, the M-step with respect to each parameter can be performed independently. (For $P$ the M-step is described in detail in the previous section.)

Estimation of $\sigma^2$ for additive Gaussian noise models such as (\ref{eqn:awgn}) is a frequently-used application of the EM algorithm, but we give the derivation here for completeness. Starting with the first term in (\ref{eqn:em-noise-derivation-1}), we can write
\begin{align}
    \sum_{k=1}^n E_{\bar{P},\bar{\sigma}^2} \Big[ \log p(y_k \given s_k; \sigma^2) \given y \Big]
    \label{eqn:em-noise-derivation-2}
    &= \sum_{k=1}^n \sum_{s_k} p(s_k \given y; \bar{P},\bar{\sigma}^2) \log p(y_k \given s_k; \sigma^2)
\end{align}
The term $p(s_k \given y; \bar{P},\bar{\sigma}^2)$ is obtained from the sum-product algorithm, and is the {\em a posteriori} probability of $s_k$ given $y$, obtained from (\ref{eqn:aposteriori}). Meanwhile, letting $I_{s_k}$ represent the current flowing through the patch clamp in each state $s_k$, from (\ref{eqn:awgn}) we have 
\begin{align}
    \log p(y_k \given s_k; \sigma^2) &= -\frac{1}{2} \log 2 \pi \sigma^2 - \frac{(y_k -I_{s_k})^2}{2\sigma^2}.
\end{align}
Let 
\begin{align}
    W &= \frac{1}{n} \sum_{k=1}^n \sum_{s_k} (y_k - I_{s_k})^2 p(s_k \given y; \bar{P},\bar{\sigma}^2) .
\end{align}
Then
(\ref{eqn:em-noise-derivation-2}) becomes
\begin{align}
    \sum_{k=1}^n E_{\bar{P},\bar{\sigma}^2} \Big[ \log p(y_k \given s_k; \sigma^2) \given y \Big]
    \label{eqn:em-noise-derivation-3}
    &= - n \frac{1}{2} \log 2 \pi \sigma^2 - n \frac{W}{2 \sigma^2}  .
\end{align}
Calculation of (\ref{eqn:em-noise-derivation-3}) and (\ref{eqn:em-derivation-3}) constitute the E-step of this EM algorithm.

In the M-step, it can be shown that the maximizing value of $\sigma^2$ in (\ref{eqn:em-noise-derivation-3}) is
\begin{align}
    \label{eqn:em-noisy-derivation-4}
    \sigma^2 = W .
\end{align}
Setting $\sigma^2$ as in (\ref{eqn:em-noisy-derivation-4}) and $P_{ij}$ as in (\ref{eqn:em-derivation-4}) completes the M-step of this EM algorithm.

\end{document}